\documentclass{article}
\usepackage{spconf,amsmath,graphicx}
\usepackage{graphicx}
\usepackage{booktabs}
\usepackage{comment}
\usepackage{amssymb}
\usepackage{todonotes}
\usepackage{multirow}

\newcommand{\fref}[1]{Figure~\ref{#1}}
\newcommand{\tref}[1]{Table~\ref{#1}}

\newcommand{\nb}[1]{\textcolor{orange}{}}

\title{Disentangled multidimensional metric learning for music similarity}
\name{Jongpil Lee$^{1*}$ \hspace{.75cm} Nicholas J. Bryan$^{2}$ \hspace{.75cm} Justin Salamon$^{2}$ \hspace{.75cm} Zeyu Jin$^{2}$ \hspace{.75cm} Juhan Nam$^{1}$\thanks{* This work was performed during an internship at Adobe Research.}}
\address{$^1$Graduate School of Culture Technology, KAIST, Daejeon, South Korea\\
$^2$Adobe Research, San Francisco, CA, USA}
\begin{document}
%\ninept
%
\maketitle
%%%%%%%%%%%%%%%%%%%%%%%%%%%%%%%%%%%%%%%%%%%%%%%%%%%%%%%%%%%%%%%%%%%%%%%%%%%%%%%%
%%%%%%%%%%%%%%%%%%%%%%%%%%%%%%%%%%%%%%%%%%%%%%%%%%%%%%%%%%%%%%%%%%%%%%%%%%%%%%%%
%%%%%%%%%%%%%%%%%%%%%%%%%%%%%%%%%%%%%%%%%%%%%%%%%%%%%%%%%%%%%%%%%%%%%%%%%%%%%%%%
\begin{abstract}
Music similarity search is useful for a variety of creative tasks such as replacing one music recording with another recording with a similar ``feel'', a common task in video editing. For this task, it is typically necessary to define a similarity metric to compare one recording to another. Music similarity, however, is hard to define and depends on multiple simultaneous notions of similarity (i.e. genre, mood, instrument, tempo). While prior work ignore this issue, we embrace this idea and introduce the concept of multidimensional similarity and unify both global and specialized similarity metrics into a single, semantically disentangled multidimensional similarity metric. To do so, we adapt a variant of deep metric learning called conditional similarity networks to the audio domain and extend it using track-based information to control the specificity of our model. We evaluate our method and show that our single, multidimensional model outperforms both specialized similarity spaces and alternative baselines. We also run a user-study and show that our approach is favored by human annotators as well.
\end{abstract} % NEED to add comment on user study...

% When text-based information is not available,

% To compute content-based music similarity typically involves extracting a feature representation from audio recordings and computing the similarity (or distance) between them using a metric or score function. 
%A common limitation of these approaches is that similarity is modeled as uni-dimensional, i.e.~songs are modelled as similar or dissimilar along a single global axis. In actuality, music is a multidimensional phenomenon, and consequently there are various \emph{different} dimensions along which songs can be compared (e.g.~timbre, rhythm, genre, mood, etc.), and songs can be simultaneously similar along some dimensions, while different along others, as illustrated in Figure \ref{fig:headline}.

%%%%%%%%%%%%%%%%%%%%%%%%%%%%%%%%%%%%%%%%%%%%%%%%%%%%%%%%%%%%%%%%%%%%%%%%%%%%%%%%
\vspace{-1mm}
\begin{keywords}
multidimensional music similarity, metric learning, disentangled representation, query-by-example.
\end{keywords}

%%%%%%%%%%%%%%%%%%%%%%%%%%%%%%%%%%%%%%%%%%%%%%%%%%%%%%%%%%%%%%%%%%%%%%%%%%%%%%%%
%%%%%%%%%%%%%%%%%%%%%%%%%%%%%%%%%%%%%%%%%%%%%%%%%%%%%%%%%%%%%%%%%%%%%%%%%%%%%%%%
%%%%%%%%%%%%%%%%%%%%%%%%%%%%%%%%%%%%%%%%%%%%%%%%%%%%%%%%%%%%%%%%%%%%%%%%%%%%%%%%
\vspace{-2mm}
\section{Introduction}
\vspace{-1mm}
\label{sec:intro}
% MOTIVATION: WHY WE NEED SEARCH BY EXAMPLE
Traditional music \emph{search} methods such as those available on streaming services and online music repositories use text-based metadata (e.g.~song, artist, album, and/or semantic tags) for music retrieval. However, there are scenarios where music metadata is either unavailable or insufficient: a concrete example is what we shall refer to as the ``music replacement'' problem, where a user wishes to replace one music recording with another recording that has a similar ``feel'', a common use case e.g.~in video editing. Describing the desired musical traits may be extremely hard to do with text, but the user has an example of what they are searching for, and so query-by-example, and more specifically content-based music similarity and retrieval, is an attractive solution. 
% WHY MUSIC RECOMMENDATION APPROACHES DON'T WORK HERE
% When text-based metadata is not available, query-by-example audio similarity-based retrieval can be employed. In this case, a query music recording is used as input to retrieve similar sounding recordings from a large  collection of music. 
While content-based music similarity has been studied extensively~\cite{casey2008content}, it has found limited application in music recommendation platforms, which rely most heavily on interaction and metadata based collaborative filtering~\cite{celma2010music}. Such techniques are not applicable, however to the music replacement scenario, where there may be little-to-no interaction data, and a user's past music replacement selections can have little correlation with future replacement needs. 
%
% Applications of audio similarity retrieval that demand highly specific notions of similarity include track identification (i.e. fingerprinting), while applications that demand low specificity include category-based retrieval~\cite{casey2008content, grosche2012audio, choi2019zero}. In the middle is the so-called semantic gap, which divides signal and semantic similarity and includes applications such as finding a similar sounding recording (``sounds like`` problem)~\cite{casey2008content}. %This problem is useful for a variety of music content-creation tasks including 1) music production for video, in which case producers will often make a ``rough cut'' film using copyrighted music and later need to replace it with similar sounding content and 2) machine assistant music composition, where finding similar music content can be used for creative inspiration.
\begin{figure}[t]
\vspace{-5mm}
\centering\includegraphics[width=.5\columnwidth, trim=0cm 1cm 0cm 0cm]{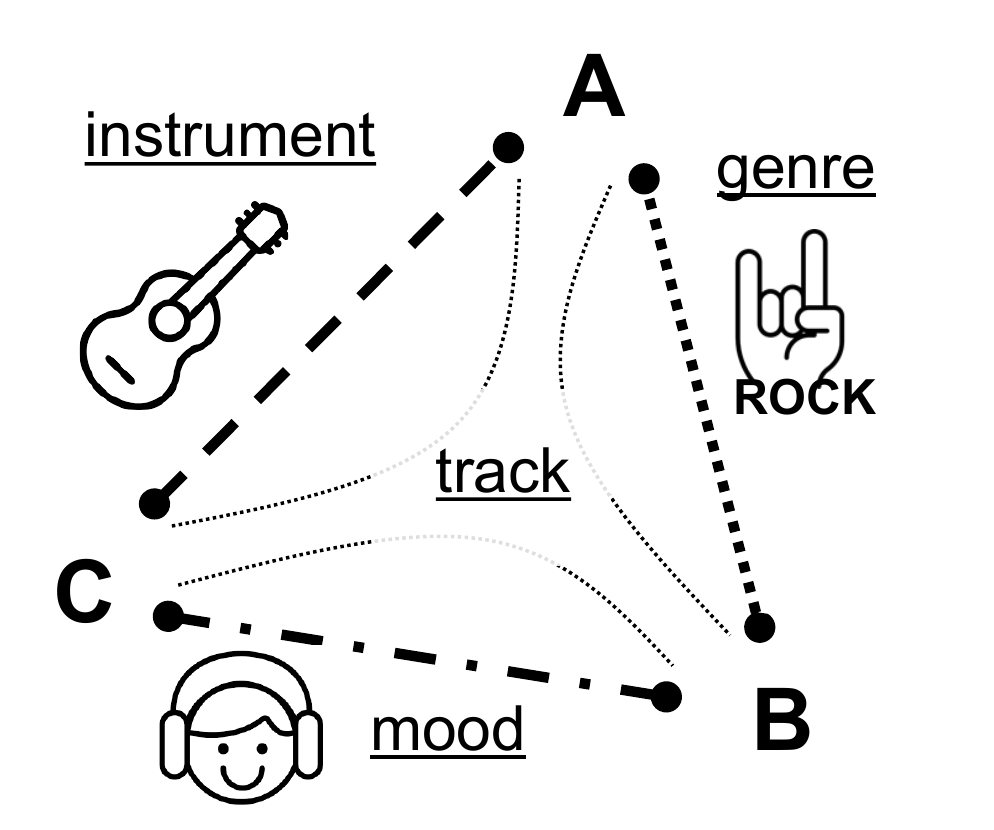}
\caption{An illustration of multiple dimensions of music similarity. Letters (A, B, C) denote different music recordings, while lines denote different dimensions of similarity. }
\vspace{-5mm}
\label{fig:headline}
\end{figure}
%
% MUSIC SPECIFICITY
From a retrieval specificity perspective, music replacement is less specific than music identification (fingerprinting), but more specific than tag-based retrieval (e.g.~genre) or than finding similar-sounding music for listening purposes \cite{casey2008content,grosche2012audio}, 
% Music replacement is typically more pragmatic (find music that can serve the same purpose as an example recording), and is typically expected to sound as close as possible to the query without being identical.
since the pragmatic goal of music replacement is to find songs which sound as close as possible to a query without being identical.

% PAST WORK IN MUSIC SIMILARITY ADN THE LACK OF MULTIDIMESIONALITY
% To find similar sounding music, similarity metrics are typically used to compute the distance between two recordings. Relevant past similarity metric methods including 
Content-based music similarity typically involves extracting a feature representation from audio recordings and computing the similarity (or distance) between them using a metric or score function. Previous approaches include vector quantization~\cite{logan2001music}, linear metric learning~\cite{mcfee2012learning, slaney2008learning, wolff2012systematic}, and, more recently, deep metric learning~\cite{wang2014learning, hoffer2015deep, jansen2018unsupervised} using human similarity labels~\cite{lu2017deep}, artist labels~\cite{park2017representation}, track labels~\cite{lee2019representation}, or tags in the context of zero-shot learning~\cite{choi2019zero}. 
% Using these approaches, however, difficulties arise because 1) music is multidimensional and multiple notions of similarity across the semantic gap inherently coexist (e.g. timbre, rhythm, genre, mood, feel, etc.), 2) different users can interpret similarity along different axes and, 3) the saliency of each similarity notion varies per recording.
A common limitation of these approaches is that similarity is modeled as uni-dimensional, i.e.~songs are modelled as similar or dissimilar along a single global axis. In actuality, music is a multidimensional phenomenon, and consequently there are various \emph{different} dimensions along which songs can be compared (e.g.~timbre, rhythm, genre, mood, etc.), and songs can be simultaneously similar along some dimensions, while different along others, as illustrated in Figure \ref{fig:headline}. It is also hard to determine precisely which dimensions people take into account when rating songs for similarity, or how they weight the importance of these dimensions. For this reason, from an application standpoint it can be beneficial to allow the user to specify which musical dimensions they care about when searching-by-example and how to weight their importance.

% WHAT WE PROPOSE - OUR CONTRIBUTIONS
% To address these issues, we propose a new deep disentangled metric learning method that bridges the semantic gap for music similarity. We do so by extending and generalizing past work~\cite{veit2017conditional} and jointly modeling multiple notions of similarity (e.g. track, genre, mood, instrument, tempo) across the specificity spectrum (low and high) within a single network using a disjoint feature embedding. Using this approach, we can model multiple individual concepts of similarity, a global notion of similarity, or any combination thereof using a single network. To evaluate our approach, we compare against past baselines and show that our method not only outperforms prior work in terms of a single global notion of similarity, but also all individual notions of similarity when compared to specialized networks. Furthermore, we collected a new dataset of human-labeled similarity ratings and show \nb{TODO}. Finally, we analyze our learned embedding features using representational similarity analysis so as to provide additional scientific understanding of our disentangled embedding.
In this paper, we propose a deep \emph{disentangled} metric learning method for learning a multidimensional music similarity space (embedding). We adapt Conditional Similarity Networks \cite{veit2017conditional}, previously only applied to images, to the audio domain, and employ a combination of user-generated tags and algorithmic estimates (i.e. tempo) to train a disentangled embedding space composed of sub-spaces corresponding to similarity along different musical dimensions: genre, mood, instrumentation and tempo. Further, we propose a track-regularization technique to increase overall perceptual similarity across all dimensions as judged by humans. We evaluate our approach against several baselines, showing our proposed approach outperforms them both in terms of global similarity and similarity along specific dimensions. To validate our quantitative results, we run a user-study and show that our proposed approach is favored by human annotators as well.

%%%%%%%%%%%%%%%%%%%%%%%%%%%%%%%%%%%%%%%%%%%%%%%%%%%%%%%%%%%%%%%%%%%%%%%%%%%%%%%%
%%%%%%%%%%%%%%%%%%%%%%%%%%%%%%%%%%%%%%%%%%%%%%%%%%%%%%%%%%%%%%%%%%%%%%%%%%%%%%%%
%%%%%%%%%%%%%%%%%%%%%%%%%%%%%%%%%%%%%%%%%%%%%%%%%%%%%%%%%%%%%%%%%%%%%%%%%%%%%%%%
\vspace{-5mm}
\section{Learning Model}

%%%%%%%%%%%%%%%%%%%%%%%%%%%%%%%%%%%%%%%%%%%%%%%%%%%%%%%%%%%%%%%%%%%%%%%%%%%%%%%%
%%%%%%%%%%%%%%%%%%%%%%%%%%%%%%%%%%%%%%%%%%%%%%%%%%%%%%%%%%%%%%%%%%%%%%%%%%%%%%%%
\vspace{-1mm}
\subsection{Metric learning with triplet loss}\label{sec:triplets}
We use deep metric learning with a triplet loss as the basis for our learning model \cite{wang2014learning, hoffer2015deep}. On a high level, our model is presented with a triplet of samples, where one is considered the ``anchor'' and the other two consist of a ``positive'' and a ``negative'', and the model is trained to map the samples into an embedding space where the ``positive'' is closer to the ``anchor'' than the ``negative'', as illustrated in \fref{fig:figure3}(A).

Formally, we define training triplets as a set $T=\{t^i\}_{i=1}^{N}$, where each triplet $t^i=\{x_a^i,x_p^i,x_n^i|s(x_a,x_p)>s(x_a,x_n)\}$, $x_a$ is the anchor sample, $x_p$ is the positive sample, $x_n$ is the negative sample, and $s$ is the musical dimension along which similarity is measured. Then, we define the triplet loss as:
\vspace{-1mm}
\begin{equation} \label{eq:eq1}
L(t)=\max\{0,D(x_a,x_p)-D(x_a,x_n)+\Delta\},
\end{equation}
where $D(x_i,x_j) = ||f(x_i)-f(x_j)||_2$ is the euclidean distance between two audio embeddings, $\Delta$ is a margin value to prevent trivial solutions, and $f(\cdot)$ is a nonlinear embedding function or deep neural network that maps the audio input to the embedding space. For a given set $T$ and embedding function $f(\cdot)$, we use stochastic gradient descent to update the network weights and minimize the loss.

%However, if we have multiple similarity notions $s$, the nonlinear function cannot take into account different similarity notions separately.  

%%%%%%%%%%%%%%%%%%%%%%%%%%%%%%%%%%%%%%%%%%%%%%%%%%%%%%%%%%%%%%%%%%%%%%%%%%%%%%%%
%%%%%%%%%%%%%%%%%%%%%%%%%%%%%%%%%%%%%%%%%%%%%%%%%%%%%%%%%%%%%%%%%%%%%%%%%%%%%%%%
\vspace{-2mm}
\subsection{Disentangling the embedding features}\label{sec:disentangling}
\begin{figure}[t!]
\vspace{-3mm}
\vspace{-4mm}
\centering\includegraphics[width=0.99\columnwidth, trim=1cm 1cm 1cm 1cm]{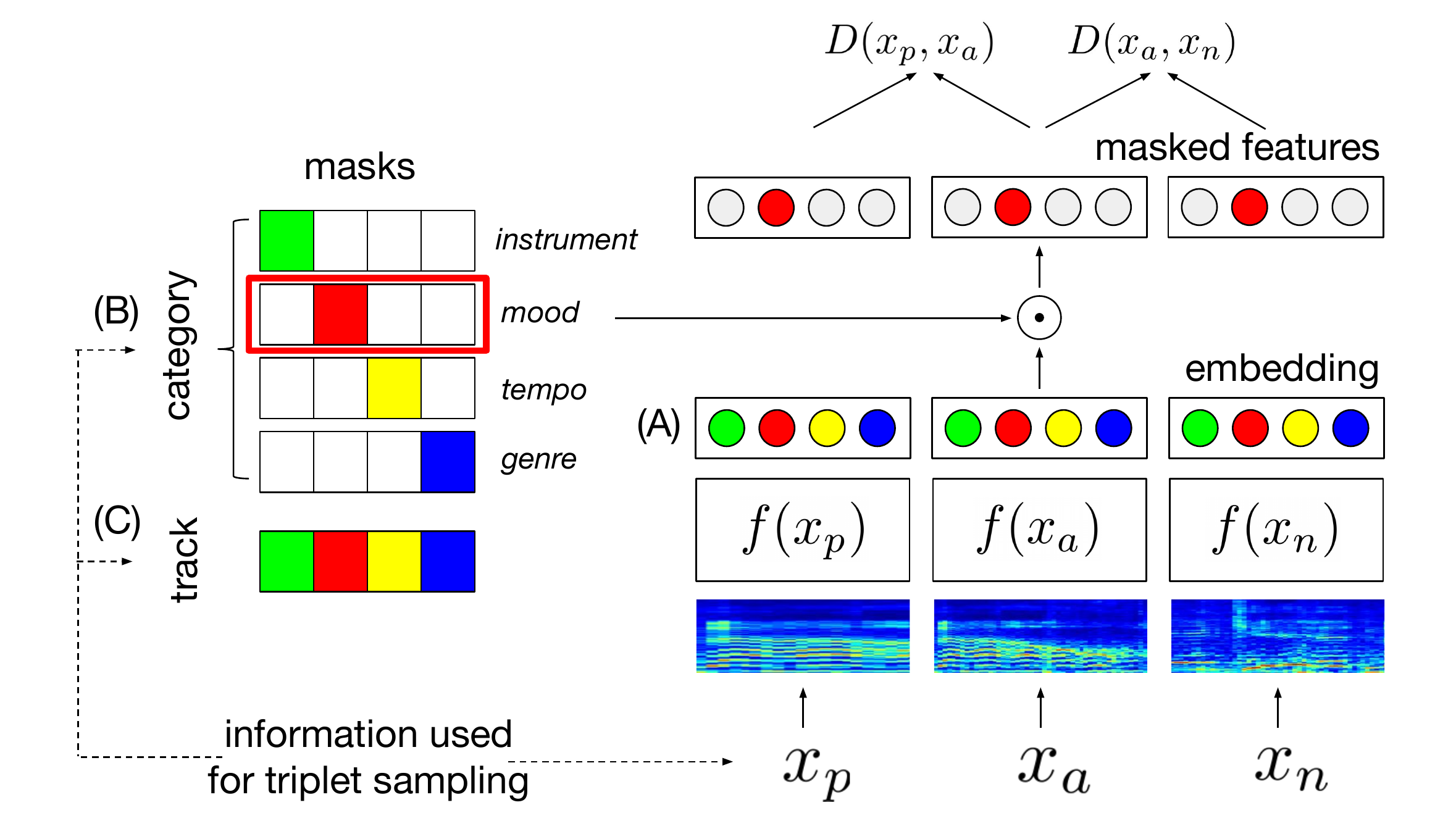}
\caption{Our proposed approach. (A) Standard triplet-based deep metric model, (B) conditional similarity masking, and (C) track regularization. }
\vspace{-2mm}
\label{fig:figure3}
\end{figure}

To jointly model multiple semantic dimensions of similarity within a single network, we adapt the work of Veit et al.~\cite{veit2017conditional}, which proposed the use of Conditional Similarity Networks (CSN)~\cite{veit2017conditional} for attribute-based image retrieval. 
% The method works by associating multiple similarity notions $s$, one-to-one, with masking functions $m_s\in \mathbb{R}^d$, where $d$ is embedding dimension. The masking function are then used to activate or block disjoint regions of the embedding features as shown in~\fref{fig:figure3}, excluding the track information of the lower left, which we address below.
The method introduces masking functions $m_s\in \mathbb{R}^d$, which are applied to the embedding space of size $d$. Each mask corresponds to a certain similarity dimension $s$ (denoted ``condition'' in \cite{veit2017conditional}), e.g.~mood or tempo, and is used to activate or block disjoint regions of the embedding space, as illustrated in~\fref{fig:figure3}(B).

Given a specific similarity dimension $s$, training triplets are defined as $T_s=\{t_s^i\}_{i=1}^{N}$, with each triplet given by:
\vspace{-1mm}
\begin{equation} \label{eq:eq2}
t_s^i=(x_a^i,x_p^i,x_n^i;s),
\end{equation}
and the training set combining triplets sampled from all similarity dimensions is defined as $T_S=\{T_s\}_{s=1}^{S}$. Consequently, we update the distance function to:
\vspace{-1mm}
\begin{equation} \label{eq:eq3}
D(x_i,x_j; s) = ||f(x_i)\circ m_s-f(x_j)\circ m_s||_2,
\end{equation}
such that the mask $m_s$ only passes through the subspace of embedding features corresponding to similarity dimension $s$ during training and $\circ$ denotes Hadamard product. Accordingly, the loss is updated to:
\vspace{-1mm}
\begin{equation} \label{eq:eq4}
L(t_s) = \max\{0,D(x_a,x_p;m_s)-D(x_a,x_n;m_s)+\Delta\}.
\end{equation}

%%%%%%%%%%%%%%%%%%%%%%%%%%%%%%%%%%%%%%%%%%%%%%%%%%%%%%%%%%%%%%%%%%%%%%%%%%%%%%%%
%%%%%%%%%%%%%%%%%%%%%%%%%%%%%%%%%%%%%%%%%%%%%%%%%%%%%%%%%%%%%%%%%%%%%%%%%%%%%%%%
\vspace{-4mm}
\subsection{Track regularization}\label{sec:regularization}
% To go beyond jointly modeling multiple disjoint semantic notions of similarity within a single network, we additional propose the use of track regularization. More specifically, we propose using track information-based triplets \cite{lee2019representation} in addition to multiple semantic category-based triplets. Track information-based triplets simply mean triplets that consist of material from two recordings from the song and a third recording a different song.  We can think of this information as akin to the fingerprinting task, which is located in high specificity spectrum of audio similarity problems.
As noted earlier, music replacement requires retrieved songs to sound as close as possible to the query example. To this end, we propose to complement the aforementioned multidimensional metric learning approach with a regularization technique we refer to as ``track regularization''. The approach involves sampling an additional set of triplets solely based on the track (song) information: the anchor and positive are both sampled from the same song, while the negative is sampled from a different song. While this sampling was used previously to learn high-specificity music similarity directly~\cite{lee2019representation}, here we use it as a ``similarity regularization'' technique to enforce a certain degree of consistency across the entire (multidimensional) embedding space. 
% Each subspace captures similarity along a certain dimension, but must also ensure that
%
% To more clearly denote the conceptual difference between semantic category-based triplets and track information-based triplet, we formulate the loss for our complete model as  
With this regularization, our final loss is given by:
\vspace{-1mm}
\begin{equation} \label{eq:eq4}
L(t_c,t_t) = L(t_c) + \lambda L(t_t),
\end{equation}
where $t_c$ are all triplets sampled from the various music similarity dimensions corresponding to disjoint sub-embedding spaces, $t_t$ are triplets sampled using track information, and $\lambda$ allows us to control the trade-off between semantic similarity (low-specificity) and overall track-based similarity (high specificity). Importantly, for track-based triplets, we use a mask with a value of one for all feature dimensions, meaning the regularization is applied to the complete embedding space to capture track similarity across all musical dimensions. Alternatively, this can be thought of as not applying any masking on the embedding space.% when computing the loss for track-based triplets.

%%%%%%%%%%%%%%%%%%%%%%%%%%%%%%%%%%%%%%%%%%%%%%%%%%%%%%%%%%%%%%%%%%%%%%%%%%%%%%%%
%%%%%%%%%%%%%%%%%%%%%%%%%%%%%%%%%%%%%%%%%%%%%%%%%%%%%%%%%%%%%%%%%%%%%%%%%%%%%%%%
%%%%%%%%%%%%%%%%%%%%%%%%%%%%%%%%%%%%%%%%%%%%%%%%%%%%%%%%%%%%%%%%%%%%%%%%%%%%%%%%
\vspace{-2mm}
\vspace{-1mm}
\vspace{-1mm}
\vspace{-1mm}
\section{Experimental Design}
% In this section, we introduce how we conduct experiments including dataset, feature extraction, evaluation method, model architecture, and comparison baseline model.

%%%%%%%%%%%%%%%%%%%%%%%%%%%%%%%%%%%%%%%%%%%%%%%%%%%%%%%%%%%%%%%%%%%%%%%%%%%%%%%%
%%%%%%%%%%%%%%%%%%%%%%%%%%%%%%%%%%%%%%%%%%%%%%%%%%%%%%%%%%%%%%%%%%%%%%%%%%%%%%%%
\vspace{-1mm}
\vspace{-1mm}
\subsection{Dataset and input features}
For our experiments, we use the Million Song Dataset (MSD)~\cite{bertin2011million}. Based on preliminary user studies on music replacement, we identify four relevant musical dimensions to consider: \textit{genre}, \textit{mood}, \textit{instrumentation}, and \textit{tempo}. To determine whether two songs are similar along these dimensions, we use Last.FM tag annotations associated with MSD tracks which have been previously grouped into different categories~\cite{choi2017convolutional}, resulting in 28 genre tags, 12 mood tags, and 5 instrument tags. Since the annotations lack tempo tags, we extract an algorithmic tempo estimate per track using the Madmom Python library~\cite{bock2015accurate,bock2016madmom}. Two tracks are considered similar along a certain musical dimension (genre, mood, instruments) if they share at least one tag in that category, or are within 5 BPM of each other in the case of tempo. For track-based triplets, we ensure there is no more than 50\% overlap between the anchor and positive samples. We split the data following~\cite{lee2017multi}, giving 201680, 11774, and 28435 samples for the train, validation, and test sets, respectively.

For training, we use 3-second excerpts represented as a log-scaled mel-spectrogram $S$, extracted with librosa~\cite{mcfee2015librosa}. We use a window size of 23 ms with 50\% overlap and compute 128 mel-bands per frame with the following log-compression: $log_{10} (1+10*S)$, resulting in input dimensions of $129\times128$ as in~\cite{park2017representation}. The representation is z-score standardized using fixed mean and standard deviation values of 0.2 and 0.25, respectively.

%%%%%%%%%%%%%%%%%%%%%%%%%%%%%%%%%%%%%%%%%%%%%%%%%%%%%%%%%%%%%%%%%%%%%%%%%%%%%%%%
%%%%%%%%%%%%%%%%%%%%%%%%%%%%%%%%%%%%%%%%%%%%%%%%%%%%%%%%%%%%%%%%%%%%%%%%%%%%%%%%

\vspace{-1mm}
\vspace{-1mm}
\vspace{-1mm}
\subsection{Model architecture and training parameters}
% Shared basic architecture for triplet network is chosen to be inception-block based model after running extensive experiments using state-of-the-art convolutional building blocks \cite{kim2019comparison}. 
For choosing the triplet network architecture, we ran preliminary experiments with several state-of-the-art convolutional building blocks~\cite{kim2019comparison}, including a basic conv-batchnorm-maxpool block, ResNet~\cite{he2016deep}, Squeeze-and-Excitation~\cite{hu2018squeeze}, and Inception~\cite{szegedy2015going}. Having identified the Inception block as the best option, we use the following model architecture: we start with 64 convolutional filters with a $5\times5$ kernel followed by $2\times2$ strided max-pooling, followed by six Inception blocks each comprising a ``na\"{i}ve'' inception module with stride $2$ followed by another inception module with a final output dimensionality of $256$ \cite{szegedy2015going}. We use ReLU nonlinearities for all layers, and apply $L2$ normalization to the embedding features prior to computing the distance~\cite{jansen2018unsupervised}.

% We tested basic conv-batchnorm-maxpool block, ResNet block, Squeeze-and-Excitation block, and Inception block. And, the final model showed overall 2\% performance increments in current evaluation setting compared to the basic block. 

% The final model first start with $64$ convolution $5\times5$ sized filters and $2\times2$ sized max-pooling. Then, six inception blocks are stacked and each inception block contains naive version inception module with stride $2$ followed by inception module with feature dimension reductions to $256$ \cite{szegedy2015going}. At last, final $256$ dimensional embedding layer is added. 
% All the layers contain ReLU nonlinearity function. We also applied $L2$ normalization on embedding features to avoid feature explosion \cite{jansen2018unsupervised}. 
% For the mask $m_s$, we explicitly divided the embedding feature dimension by the number of similarity notions that are used to construct sub-embedding spaces and assigned ones separately for each mask. In this paper, we used 4 similarity notions, so each $m_s$ contains 64 values of one and others are zeros. For example, the first 64 dimension has a value of one and others are zeros for \textit{genre} mask. 

Since our total embedding size is 256 and we consider four music similarity dimensions (genre, mood, instruments, tempo), each with a disjoint subspace of size 64. We also experimented with a trainable masking layer~\cite{veit2017conditional} (as opposed to fixed disjoint masks), but found it did not lead to any significant improvement. Moreover, using fixed masks has the added benefit of allowing us to weight each musical dimension independently post-hoc which, as noted earlier, is a desirable user interaction paradigm.
%
% We also tested trainable masking layer as presented in the CSN paper \cite{veit2017conditional}, but the results did not give meaningful difference compared to the disjoint mask, so we only included the results of disjoint masks for simplification. 
%
% All training is performed using Adam optimizer, and learning rate was set to $0.01$ with early stopping strategy. The learning rate was reduced 5 times by factor of 5 when the validation loss does not decrease after 4 epochs from the least validation loss. The margin value of triplet loss is set to $0.1$ in all cases. The $\lambda$ value is set to 0.5 when track regularization is applied.
%
We use the Adam optimizer~\cite{kingma2014adam} for training. We initialize the learning rate to $0.01$ and reduce it by a factor of 5 when the validation loss does not decrease for 4 epochs, up to 5 times, after which we apply early stopping. The margin for the triplet loss is set to $0.1$. And, after empirically hearing the properties of similarity space, $\lambda$ was set to 0.5 when track regularization is applied.
%%%%%%%%%%%%%%%%%%%%%%%%%%%%%%%%%%%%%%%%%%%%%%%%%%%%%%%%%%%%%%%%%%%%%%%%%%%%%%%%
%%%%%%%%%%%%%%%%%%%%%%%%%%%%%%%%%%%%%%%%%%%%%%%%%%%%%%%%%%%%%%%%%%%%%%%%%%%%%%%%
\vspace{-1mm}
\vspace{-1mm}
\vspace{-1mm}
\vspace{-1mm}
\vspace{-1mm}
\vspace{-1mm}
\subsection{Evaluation metrics and user-study}

\begin{table*}[ht]
\vspace{-1mm}
\vspace{-1mm}
\vspace{-7mm}
\centering
\resizebox{0.77\textwidth}{!}{\begin{tabular}{@{}ccccccc@{}}
\toprule
Used space & Embedding Features & Genre & Mood & Instruments & Tempo & Overall \\ \midrule
\multirow{6}{*}{All-dimensions} & MFCC-VQ & 0.563 & 0.481 & 0.495 & 0.516 & 0.514 \\
 & Track & 0.611 & 0.595 & 0.531 & 0.534 & 0.568 \\
 & Category & 0.647 & 0.633 & 0.562 & 0.875 & 0.679 \\
 & Category + track regularization & 0.647 & 0.627 & 0.561 & 0.891 & 0.681 \\
 & Category + disentanglement & 0.708 & 0.717 & 0.657 & 0.783 & 0.716 \\
 & Category + disentanglement + track regularization & 0.693 & 0.704 & 0.626 & 0.836 & 0.715 \\ \midrule
\multirow{3}{*}{Sub-dimensions} & Set of specialized networks & 0.708 & 0.619 & 0.603 & 0.942 & 0.718 \\
 & Category + disentanglement & 0.785 & 0.790 & 0.798 & 0.955 & 0.832 \\
 & Category + disentanglement + track regularization & 0.765 & 0.743 & 0.700 & 0.953 & 0.790 \\ \bottomrule
\end{tabular}}
\caption{Prediction accuracy of category-based (genre, mood, instruments, tempo) triplets.}
\vspace{-3mm}
\label{table:table1}
\end{table*}

\begin{table}[ht]
\vspace{-0mm}
\centering
\resizebox{0.9\columnwidth}{!}{\begin{tabular}{@{}ccc@{}}
\toprule
Embedding Features                          & Track & User \\ \midrule
MFCC-VQ                        & 0.833 & 0.654                        \\
Track                       & 0.950 & 0.763                      \\
Category                    & 0.975 & 0.766                      \\
Category + track regularization            & 0.980 & 0.740                      \\
Category + disentanglement         & 0.985 & 0.763                      \\
Category + disentanglement + track regularization & 0.988 & 0.792                      \\ \bottomrule
\end{tabular}}
\caption{Results on track-based and user-based triplets.}
\vspace{-3mm}
\label{table:table2}
\end{table}

% All the evaluations are conducted in held-out triplet prediction setting. We first built 5 held-out triplet sets using following similarity notions : \textit{genre}, \textit{mood}, \textit{instr}, \textit{tempo}, and \textit{track}. Each contains 40,000 triplets. For the category information prediction (which are \textit{genre}, \textit{mood}, \textit{instr}, and \textit{tempo}), song-level evaluation is conducted. In this case, all the non-overlapped embedding features that are extracted from each song are averaged to represent song-level features. For the \textit{track}-based triplet prediction, we perform segment-level evaluation. This is because that the anchor and positive segments of this triplets are actually the same track.
For evaluation we use a set of held-out triplets sampled from the test set. We sample 40,000 triplets per music dimension (genre, mood, instruments, tempo) as well as 40,000 triplets based on track information. To simulate our application scenario, we use triplets of full songs for evaluation, the only exception being track-based triplets, where we stick to 3 second excerpts since the anchor and positive are sampled from the same song and should not overlap by more than 50\%. The embedding for a full song is obtained by computing embedding frames from 3-second non-overlapping windows and averaging them over the time dimension. Given a test triplet, a model is evaluated by testing whether the embedding distance between the anchor and positive samples is smaller than the distance between the anchor and negative (score of 1), or greater (score of 0). The scores for all triplets are averaged to obtain a final score between 0 (worst) and 1 (best).

%Also, we ran user experiment to acquire user-rated held-out triplets to measure how the model performs on similarity concept of user's perspectives. To do that, we randomly chose 12,000 segments (4,000 triplets) from the test tracks and give a user to rate which one is more similar to the anchor song. Each triplet is exposed to max 12 ratings. We then left triplets that showed above 0.9 agreements which result in 879 user-rated triplets. 
To determine whether human subjects concur with the above quantitative evaluation, we also randomly sampled 4,000 triplets from the test set and asked people to annotate which track sounded more similar to the anchor (positive or negative) without showing which was which. Each triplet was annotated by 5-12 people, resulting in 39,440 human annotations. We then calculated the annotator agreement per triplet, defined as the ratio between the majority vote and total number of annotations, and filtered out triplets where the agreement was below 0.9, resulting in 879 high-agreement human-annotated  triplets. Since similarity judgements have a high degree of subjectivity, in this way, we can limit the scope of our human evaluation to triplets where there is broad annotator agreement. Models are evaluated against these triplets as described earlier, obtaining a score between 0--1 in terms of consistency with user ratings. For reproducibility, we share our dataset of user similarity ratings, \textit{dim-sim}, online, along with audio similarity examples for the proposed approach \footnote{https://jongpillee.github.io/multi-dim-music-sim/}.

%%%%%%%%%%%%%%%%%%%%%%%%%%%%%%%%%%%%%%%%%%%%%%%%%%%%%%%%%%%%%%%%%%%%%%%%%%%%%%%%
%%%%%%%%%%%%%%%%%%%%%%%%%%%%%%%%%%%%%%%%%%%%%%%%%%%%%%%%%%%%%%%%%%%%%%%%%%%%%%%%
\vspace{-1mm}
\subsection{Baseline method}
%We also implemented comparison embedding features to measure baseline performance in a given data and evaluation setup. We chose to use vector quantization (VQ) approach since it showed powerful baseline in similarity-based music search and music auto-tagging \cite{mcfee2012learning,liang2014codebook}. We first run the K-means clustering algorithm on a randomly selected 2,500,000 frames of the normalized MFCCS on the training tracks. The number of cluser is set to 1024. In the evaluation procedure, all the frames in a track is assigned to the closest cluster in euclidean distance metric and aggregated in a histogram. After then, the histogram is normalized by the number of frames in a track so that the different length of audio can be controlled. We then use euclidean distance as a distance metric when comparing two histograms \cite{mcfee2012learning}.
As a strong baseline, we implement a vector quantization method that has been used for both similarity-based music retrieval and auto-tagging \cite{mcfee2012learning,liang2014codebook}. We compute 13 MFCC coefficients and their first and second derivatives per frame for each track, randomly select 2,500,000 frames from all tracks and cluster them using K-means with $K=1024$ to produce a dictionary \cite{mcfee2012learning}. Given the dictionary, a track embedding is obtained by assigning each MFCC frame to its closest cluster and computing a normalized histogram of cluster assignments. The distance between any two tracks is then given by the Euclidean distance between their normalized histograms~\cite{mcfee2012learning}.
%Given a track, we extract 13 MFCC coefficients per-frame and compute summary statistics (mean and standard deviation) for each coefficient as well as its first and second derivatives, resulting in a 39-dimensional feature vector \cite{}.

% "We compute summary statistics (mean and standard deviation) for each MFCC dimensions as well as its first and second derivatives, resulting in a 39-dimensional feature vector [5]"

%%%%%%%%%%%%%%%%%%%%%%%%%%%%%%%%%%%%%%%%%%%%%%%%%%%%%%%%%%%%%%%%%%%%%%%%%%%%%%%%
%%%%%%%%%%%%%%%%%%%%%%%%%%%%%%%%%%%%%%%%%%%%%%%%%%%%%%%%%%%%%%%%%%%%%%%%%%%%%%%%
%%%%%%%%%%%%%%%%%%%%%%%%%%%%%%%%%%%%%%%%%%%%%%%%%%%%%%%%%%%%%%%%%%%%%%%%%%%%%%%%
\vspace{-6mm}
\section{Results}
%Here, we verify and analyze the disentangled embedding space. We first report the results on held-out triplet prediction tasks and then analyze how each of the disentangled embedding feature spaces are related.
\vspace{-1mm}

% \subsection{Triplet prediction}
In Table~\ref{table:table1}, we present the numerical results obtained for each of the four held-out triplet sets corresponding to a music similarity dimension, as well as aggregated scores over all four triplet sets (``Overall'').
%Used space indicates that whether to use all-dimensions of embedding features for comparing pair of data, so sub-dimension means that embedding features are masked with the mask of evaluating similarity notion in the evaluation. We compared six models including baseline MFCC+VQ. ''Category'' and ''Track'' indicates that which set of triplets are used in training (category set contains \textit{genre}, \textit{mood}, \textit{instr}, and \textit{tempo}) and ''with Disentanglement'' refers to the use of disentangling embedding features. Also, the ''Set of Specialized Networks'' means that 4 models each are trained with target similarity notions \cite{veit2017conditional} are used in evaluation of each similarity notion. From the results, we can first find that disentangling technique improves the performance a lot in both all-dimensions and sub-dimension settings. However, when we compare ''Category with Disentanglement'' with ''Category with Disentanglement + Track Regularization'', it seems like adding track regularization harms the individual sub-dimension's performance on each similarity notion. This makes sense that track regularization is actually regularizing sub-dimensions while maintaining overall embedding characteristics. 
The ``Used space'' column indicates which subset of the embedding space was used to compute the distance between pairs of tracks, where ``all dimensions'' means all embedding features were used ($f(x)$), whereas ``sub-dimensions'' means only the subspace corresponding to the musical dimension ($f(x)\circ m_s$) from which the test triplets were sampled was used. We compare six models plus the baseline, specified in the ``embedding features'' column. The ``Track'' model was trained on triplets sampled based on track-information only, the ``Category'' model was trained on triplets sampled from the four music similarity dimensions (categories) of genre, mood, instruments and tempo, including both with and without disentanglement (subspace masking) and track regularization. For disentangled models, we include an additional baseline, ``Set of specialized networks'', which is comprised of four separate triplet-loss networks, each trained exclusively on triplets sampled from one of the four musical dimensions.

We see that all deep metric learning models outperform the MFCC-VQ baseline. More importantly, disentangling the embedding improves performance in almost all cases, with our disentangled model trained on all triplets jointly (Category + disentanglement) even outperforming the specialized networks trained separately on each dimension. 

As one might expect, track regularization decreases numerical performance on each of the four triplet test sets, as it enforces all embedding subspaces to respect a global notion of track similarity. The key question is how does it affect model performance when compared against the human ratings obtained from our user study, presented in \tref{table:table2}. As a sanity check, we start by evaluating our models against the track-based triplet test-set, presented in the ``Track'' column. We see that, as expected, track-regularization increases performance on this high-specificity set. Somewhat surprisingly, training on category-sampled triplets outperforms training on track-sampled triplets, with disentanglement increasing performance further. Next, we turn to the results obtained from the user study, presented in the ``User'' column. We see that our proposed approach outperforms the baseline, and, as per our initial hypothesis, track regularization increases the overall user agreement with our model's similarity ratings when training on category triplets with disentanglement.

\vspace{-3mm}
\section{Conclusion}
\vspace{-1mm}
%In this work, we extended CSNs by adding track regularization technique to take into account both diverse types of individual similarity and global similarity. The evaluations in three different types of notions of similarity: music semantics information, track identification information, and newly created human-labeled similarity ratings, showed that the proposed model is superior to all other comparisons in all similarity notions. We can interpret this as the model can actually bridge the semantic gap between similarity notions with quite different properties. Furthermore, learning a semantically disentangled similarity spaces enables new query-by-example interaction paradigms in music similarity-based retrieval.
In this paper, we introduce a novel approach for deep metric learning of a disentangled, \emph{multidimensional}, music similarity space. We use Conditional Similarity Networks trained on a combination of user tags and algorithmic estimates, and introduce track regularization to control for retrieval specificity. Through a series of experiments, including both a quantitative evaluation and a user study, we demonstrate that our proposed approach outperforms several baselines, with per-dimension similarity performance increasing due to the disentangling of the embedding space, and agreement with human annotations increasing as a result of track regularization. Our solution is particularly relevant to the music replacement problem, and opens the door to novel interaction paradigms which permit the user to select which music dimensions they care about for retrieval, how to weight their relative importance, and how to balance subspace similarity versus high-specificity overall similarity. This approach can further be extended to general audio similarity such as voice similarity based on their speaker's condition, phonation, or prosody.
In the future, we plan to conduct further user studies to determine human agreement when considering each musical dimension in isolation, and evaluate the performance of our model against these ratings. We also plan to explore and evaluate our proposed approach for multi-query retrieval (query-by-multiple-examples) and mix-and-match scenarios where the user is interested in finding songs whose characteristics match the subspaces of different songs (e.g.~the genre of example A with the tempo of example B).

%{\footnotesize
\bibliographystyle{IEEEbib} 
\bibliography{strings,refs}
%}

\end{document}